\documentclass[10pt, prb, aps, twocolumn, showpacs, citeautoscript, floatfix, reprint, amsmath, amssymb, notitlepage, superscriptaddress, longbibliography]{revtex4-1}

\usepackage{graphicx}
\usepackage{stmaryrd}
\usepackage{rotating}
\usepackage{amsmath}
\usepackage{amsfonts}
\usepackage{amssymb}
\usepackage{wasysym}
\usepackage[countmax]{subfloat}
\usepackage{dcolumn} 
\usepackage{bm} 
\usepackage{color}

\usepackage{float}
\usepackage{latexsym}

\begin{document}

\title{Quantum geometric origin of the Meissner effect and superfluid weight marker}

\author{David Porlles}
\affiliation{Department of Physics, PUC-Rio, 22451-900 Rio de Janeiro, Brazil}

\author{Wei Chen}
\affiliation{Department of Physics, PUC-Rio, 22451-900 Rio de Janeiro, Brazil}

\date{\rm\today}

\begin{abstract}

The momentum space of conventional superconductors is recently recognized to
possess a quantum metric defined from the overlap of filled quasihole 
states at neighboring momenta. For multiband superconductors with arbitrary intraband and interband $s$-wave pairing, we elaborate that their superfluid weight in London equations is given by the momentum integration of the elements of quantum metric times the quasiparticle energy, indicating the quantum geometric origins of Meissner effect and vortex state. The momentum integration of the quantum metric further yields a spread of quasihole Wannier functions that characterizes the stability of the superconducting state. Our formalism allows the diamagnetic response of conventional superconductors to be mapped to individual lattice sites as a superfluid weight marker, which can incorporate the effect of disorder through self-consistently solving the Bogoliubov-de Gennes equations. Using single-band $s$-wave superconductors in 2D and 3D as examples, our marker reveals a diamagnetic current that becomes turbulent in the presence of nonmagnetic impurities, and the increase of London penetration depth by disorder that is consistent with experiments.

\end{abstract}

\maketitle

\section{Introduction} 

The celebrated Meissner effect\cite{Meissner33}, described phenomenologically by London equations\cite{London35}, is without a doubt one of the most significant properties of superconductors (SCs) that has countless technological applications. The effect is attributed to the occurrence of a diamagnetic current that shields the external magnetic field, which is usually larger than the accompanied paramagnetic current. Despite being well-understood for decades, the interest on this diamagnetic response resurges recently due to the interesting results obtained on the flat band materials. It is found that if the normal state band structure contains flat bands, the superfluid weight that characterizes the diamagnetic response is determined by the quantum metric of the flat bands in the normal state, manifesting another important physical phenomenon belonging to the surging notion of quantum geometry\cite{Peotta15,Julku16,Liang17,HerzogArbeitman22,Torma22,Iskin23}. To be more specific, in these materials one may separate the superfluid weight into the so-called conventional and geometric components. It is argued that because the conventional component is given by the group velocity and hence vanishes in the flat band materials, the geometric component originated from the normal state quantum metric dominants. Furthermore, in multiband SCs, the normal state quantum metric also enters the Cooper pair when it carries a finite momentum, rendering what has been called the quantum-geometric pair potential contribution to the superfluid weight\cite{Daido24}. Recent developments also suggest that quantum geometry induces ferromagnetic fluctuation and can induce spin-triplet superconductivity\cite{Kitamura24}, as well as promotes pairing from Coulomb repulsion via the Kohn-Luttinger mechanism due to the dependence of screening on the quantum metric\cite{Shavit25}, indicating more exotic types of pairing mechanisms can be the consequences of the normal state quantum metric.  



In contrast to these previous works valid only for flat band materials, our aim in this paper is to elaborate that the diamagnetic current in the Meissner effect is in fact of quantum geometric origin in any $s$-wave SCs. The quantum metric that contributes ubiquitously to the diamagnetic superfluid weight $D_{\mu\nu}^{\,d}$ is not the quantum metric of the normal state, but the {\it quasihole quantum metric} of the superconducting state defined from the overlap of fully antisymmetric quasihole states at neighboring momenta\cite{Porlles23_SC_metric}. For generic multiband SCs with any intraband and interband $s$-wave pairing, which describe the vast majority of conventional SCs of any band structure, we explicitly demonstrate the interpretation of $D_{\mu\nu}^{\,d}$ in terms of the quasihole quantum metric, thus clarifying the quantum geometric origins of the Meissner effect and the Ginzburg-Landau parameter for the occurrence of the vortex state\cite{Ginzburg50}. This quasihole quantum metric exists even if the normal state does not contain any quantum metric, and thus is a generic feature that ubiquitously manifests in any SCs. Furthermore, we point out that because the quasihole state is also a Bloch state, its Fourier transform yields a quasihole Wannier function that is generally localized around the designated unit cell. We then proceed to introduce a spread of quasihole Wannier function in a way completely analogous to that proposed for semiconductors and insulators, whose gauge-invariant part is given by the momentum integration of the quasihole quantum metric\cite{Marzari97,Souza08,Marzari12}. This spread is proposed to characterize the stability of the superconducting state, and is estimated to be determined by the coherence length and Fermi momentum.

Based on the aforementioned quantum geometric interpretation of Meissner effect, we proceed to map the superfluid weight to real space by using the formalism of topological markers\cite{Bianco11,Prodan10,Prodan10_2,Loring10}, yielding what we call the superfluid weight marker. This marker enables the calculation of local diamagnetic current on each lattice sites through self-consistently solving the Bogoliubov-de Gennes (BdG) equations of a lattice Hamiltonian, which can incorporate the effect of disorder. Using the simplest mean field $s$-wave SCs as examples, the superfluid weight marker reveals a turbulent and suppressed diamagnetic current caused by nonmagnetic impurities, offering an explanation to experimental result of the increased penetration depth by disorder first revealed by Pippard\cite{Pippard53}.

\section{Quantum geometry and Meissner effect \label{sec:quantum_geometry_Meissner}} 

\subsection{Quasihole quantum metric of conventional multiband superconductors \label{sec:quantum_metric_multiband}}

We start by formulating the quantum metric for a $D$-dimensional multiorbital conventional SC with arbitrary number of bands, and any intraband and interband $s$-wave pairing treated within mean field theory. The formalism is a straightforward generalization of the quantum metric in single-band SCs\cite{Porlles23_SC_metric}. The Nambu basis of the Hamiltonian at momentum ${\bf k}$ (with unit $[{\bf k}]=$kgm/s) is $(c_{{\bf k\uparrow}1},c_{{\bf k\uparrow}2}...c_{{\bf -k\downarrow}1}^{\dag},c_{{\bf -k\downarrow}2}^{\dag}...)$, where the numbers in the subscript enumerate the band index and the arrows the spin up or down. We collectively label all the degrees of freedom by $\gamma=\left\{e\uparrow,h\downarrow\right\}\otimes$ orbitals. After the mean field Hamiltonian $H({\bf k})$ in momentum space is diagonalized, one obtains $N_{-}$ number of filled quasihole eigenstate $|n({\bf k})\rangle\equiv|n\rangle$ with negative eigenenergy $\varepsilon_{n}<0$ and the same number of empty quasiparticle eigenstate $|m({\bf k})\rangle\equiv|m\rangle$ with positive eigenenergy $\varepsilon_{m}>0$. We consider the fully antisymmetric quasihole state that respects Fermi statistics\cite{Matsuura10,vonGersdorff21_metric_curvature} 
\begin{eqnarray}
|u^{\rm h}({\bf k})\rangle=\frac{1}{\sqrt{N_{-}!}}\epsilon^{n_{1}n_{2}...n_{N-}}|n_{1}^{\bf k}\rangle|n_{2}^{\bf k}\rangle...|n_{N_{-}}^{\bf k}\rangle.\;\;\;
\label{psi_val}
\end{eqnarray}
The quasihole quantum metric is defined from the overlap of two such states at slightly different momenta\cite{Provost80} 
\begin{eqnarray}
|\langle u^{\rm h}({\bf k})|u^{\rm h}({\bf k+\delta k})\rangle|=1-\frac{1}{2}g_{\mu\nu}({\bf k})\delta k^{\mu}\delta k^{\nu},
\label{uval_gmunu}
\end{eqnarray}
which amounts to the expression\cite{vonGersdorff21_metric_curvature} 
\begin{eqnarray}
&&g_{\mu\nu}({\bf k})=\sum_{nm}g_{\mu\nu}^{nm},
\nonumber \\
&&g_{\mu\nu}^{nm}=\frac{1}{2}\left[\langle \partial_{\mu}n|m\rangle\langle m|\partial_{\nu}n\rangle+\langle \partial_{\nu}n|m\rangle\langle m|\partial_{\mu}n\rangle\right],
\label{gmunu_T0}
\end{eqnarray}
where we have labeled each term in the summation over the quasiholes $\sum_{n}$ and quasiparticles $\sum_{m}$ by $g_{\mu\nu}^{nm}$. From this quantum metric, we can introduce yet another geometrical quantity of particular interest, namely the quantum metric integrated over the $D$-dimensional BZ
\begin{eqnarray}
&&{\cal G}_{\mu\nu}=\int\frac{d^{D}{\bf k}}{(2\pi)^{D}}g_{\mu\nu}({\bf k}),
\label{Gmunu_definition}
\end{eqnarray}
which we call the fidelity number\cite{deSousa23_fidelity_marker}. Physically, it represents the average distance between neighboring quasihole states $|u^{\rm h}({\bf k})\rangle$ and $|u^{\rm h}({\bf k+\delta k})\rangle$, and hence serves as a characteristic quantum geometrical property of the compact Euclidean manifold of BZ.

\subsection{Spread of quasihole Wannier function}

In this section, we elaborate that the trace of the fidelity number matrix in Eq.~(\ref{Gmunu_definition}) gives the gauge-invariant part of what we call the spread of quasihole Wannier functions, a concept similar to that in semiconductors and insulators\cite{Marzari97,Souza08,Marzari12}. For any of the quasiparticle $\ell=m$ or quasihole $\ell=n$ states in a multiband SC, we denote $\langle{\bf r}|\ell({\bf k})\rangle=\ell_{\bf k}({\bf r})=e^{-i{\bf k\cdot r}/\hbar}\psi_{\ell}^{\bf k}({\bf r})$ as the periodic part of the Bloch state satisfying $\ell_{\bf k}({\bf r})=\ell_{\bf k}({\bf r+R})$, where ${\bf r}$ and ${\bf R}$ are Bravais lattice vectors. From these Bloch states, one can introduce Wannier state $|{\bf R}\ell\rangle$ via a Fourier transform
\begin{eqnarray}
&&|\ell({\bf k})\rangle=\sum_{{\bf R}}e^{-i {\bf k}\cdot({\hat{\bf r}}-{\bf R})/\hbar}|{\bf R}\ell\rangle,
\nonumber \\
&&|{\bf R} \ell\rangle=\sum_{\bf k}e^{i {\bf k}\cdot({\hat{\bf r}}-{\bf R})/\hbar}|\ell({\bf k})\rangle.\;\;\;\;\;
\label{Wannier_basis}
\end{eqnarray}
We should call the Wannier function of the quasihole state $\ell=n$ the quasihole Wannier function $\langle {\bf r}|{\bf R} n\rangle=W_{n}({\bf r}-{\bf R})$, which is centering around the home cell ${\bf R}$.

We proceed to introduce the spread of quasihole Wannier function by drawing analogy with that in the semiconductors and insulators, defined by\cite{Marzari97,Marzari12}
\begin{eqnarray}
&&\Omega=\sum_{n}\left[\langle r^{2}\rangle_{n}-{\hat{\bf r}}_{n}^{2}\right]
\nonumber \\
&&=\sum_{n}\left[\langle{\bf 0}n|r^{2}|{\bf 0}n\rangle-\langle{\bf 0}n|{\bf r}|{\bf 0}n\rangle^{2}\right]=\Omega_{I}+\tilde{\Omega}.
\label{Omega_original}
\end{eqnarray}
Despite a definition completely analogous to that in semiconductors and insulators, it should be emphasized that this spread does not have the meaning of the variance of a charge distribution. Rather, because the Nambu basis of conventional multiband SCs is a mixture of spin up electrons and spin down holes according to Sec.~\ref{sec:quantum_metric_multiband}, the spread represents the variance of the quasihole wave function that has both particle and hole components. This spread can be further separated into the gauge invariant part $\Omega_{I}$ and the gauge-dependent part $\tilde{\Omega}$ 
\begin{eqnarray}
&&\Omega_{I}=\sum_{n}\left[\langle{\bf 0}n|r^{2}|{\bf 0}n\rangle-\sum_{{\bf R}n'}|\langle{\bf R}n'|{\bf r}|{\bf 0}n\rangle|^{2}\right],
\nonumber \\
&&\tilde{\Omega}=\sum_{n}\sum_{{\bf R}n'\neq{\bf 0}n}|\langle{\bf R}n'|{\bf r}|{\bf 0}n\rangle|^{2}.
\label{Omega_gauge_inv}
\end{eqnarray}
In superconductors and insulators, it has been shown that the gauge-invariant part $\Omega_{I}$ is given by the fidelity number in Eq.~(\ref{Gmunu_definition}). This conclusion holds true for singlet SCs, as can be seen by considering the identities
\begin{eqnarray}
&&\langle r^{2}\rangle_{n}=\frac{V_{\rm cell}}{\hbar^{D-2}}\int\frac{d^{D}{\bf k}}{(2\pi)^{D}}\sum_{\mu}\langle\partial_{\nu}n|\partial_{\nu}n\rangle,
\nonumber \\
&&\langle{\bf R}n'|{\hat\mu}|{\bf 0}n\rangle=\frac{V_{\rm cell}}{\hbar^{D-1}}\int\frac{d^{D}{\bf k}}{(2\pi)^{D}}\langle n'|i\partial_{\mu}|n\rangle e^{i{\bf k\cdot R}/\hbar},
\end{eqnarray}
from which $\Omega_{I}$ can be written as\cite{Marzari97,Souza08,Marzari12} 
\begin{eqnarray}
&&\Omega_{I}=\frac{V_{\rm cell}}{\hbar^{D-2}}\int\frac{d^{D}{\bf k}}{(2\pi)^{D}}
\nonumber \\
&&\times\sum_{\mu}\sum_{n}\left[\langle\partial_{\nu}n|\partial_{\nu}n\rangle-\sum_{n'}\langle\partial_{\nu}n|n'\rangle\langle n'|\partial_{\nu}n\rangle\right]
\nonumber \\
&&=\frac{V_{\rm cell}}{\hbar^{D-2}}\,{\rm Tr}\,{\cal G}_{\mu\nu},
\label{OmegaI_trace_Gmunu}
\end{eqnarray}
where ${\rm Tr}\,{\cal G}_{\mu\nu}=\sum_{\mu\mu}{\cal G}_{\mu\mu}$. Thus the gauge-invariant part of the spread $\Omega_{I}$ is equivalently the trace of fidelity number in a $D$-dimensional conventional SC. Note that the fidelity number ${\cal G}_{\mu\nu}$ and hence the spread $\Omega_{I}$ in semiconductors and insulators can be measured by an optical sum rule of dielectric function in 3D, and analogously a sum rule of absorbance in 2D, owing to the equivalence between quantum metric and optical transition matrix element\cite{Cardenas24_spread_Wannier}. However, in conventional SCs, because the optical transition matrix elements are not exactly the quantum metric\cite{Porlles23_SC_metric}, it remains unclear to us at present whether ${\cal G}_{\mu\nu}$ and $\Omega_{I}$ can be measured experimentally by certain means.

The dependence of the spread $\Omega_{I}$ on the superconducting gap $\Delta$ can be qualitatively understood by the following analysis. For a parabolic band $\varepsilon_{\bf k}=k^{2}/2m-k_{F}^{2}/2m$ with Fermi momentum $k_{F}$ and a heuristic lattice constant $a$ introduced to regularize the integral, it has been shown that the diagonal elements of fidelity number in Eq.~(\ref{Gmunu_definition}) have the following analytical expressions in 2D and 3D
\begin{eqnarray}
&&{\cal G}_{\mu\mu}^{3D}=\frac{\pi^{2}}{6\sqrt{2}}\left(\frac{\xi}{a}\right)\left(\frac{k_{F}}{2\pi\hbar/a}\right)^{2}\left(\frac{\hbar}{a}\right),
\nonumber \\
&&{\cal G}_{\mu\mu}^{2D}
=\frac{\pi^{2}}{8\sqrt{2}}\left(\frac{\xi}{a}\right)\left(\frac{k_{F}}{2\pi\hbar/a}\right),
\end{eqnarray}
where $\xi=\hbar k_{F}/\pi m\Delta$ is the Bardeen-Cooper-Schrieffer (BCS) coherence length. Together with the relation between $\Omega_{I}$ and ${\cal G}_{\mu\mu}$ in Eq.~(\ref{OmegaI_trace_Gmunu}), we see that $\Omega_{I}$ is inversely proportional to the gap
\begin{eqnarray}
\Omega_{I}\propto{\cal G}_{\mu\mu}\propto\xi\propto\frac{1}{\Delta}.
\end{eqnarray}
This result coincides with the intuitive picture that a larger superconducting gap implies a more localized quasihole Wannier function that is harder to excite, and hence a more stable SC. Although more complicated band structures can alter this simple estimation, we anticipate that the reduction of $\Omega_{I}$ upon increasing $\Delta$ should be a generic feature for $s$-wave SCs.




\subsection{Quantum geometric origin of Meissner effect and vortex state \label{sec:origin_Meissner_effect}} 

We now elaborate that the diamagnetic current in conventional SCs is given by the quasihole quantum metric. This is done by first expanding the Hamiltonian in the presence of the vector potential ${\bf A}$ to second order\cite{Scalapino92,Scalapino93,Liang17,Rossi21} (repeated Greek indices are summed)
\begin{eqnarray}
H=H_{0}+e{\hat v}_{\mu}A^{\mu}+\frac{1}{2}V_{\rm cell}{\hat D}_{\mu\nu}^{\,d}A^{\mu}A^{\nu},
\label{Hamiltonian_expansion_vector_potential}
\end{eqnarray}
where ${\hat v}_{\mu}$ is the velocity operator that eventually leads to a paramagnetic superfluid weight ${\hat D}_{\mu\nu}^{\,p}$, which has been addressed in detail and is not our main concern in the present work. Instead, we focus on the term that is second order in the vector potential described by the diamagnetic superfluid weight operator
\begin{eqnarray}
{\hat D}_{\mu\nu}^{\,d}=\frac{e^{2}}{V_{\rm cell}}\sum_{\bf k\gamma}c_{\bf k\gamma}^{\dag}\partial_{\mu}\partial_{\nu}H_{\gamma}c_{\bf k\gamma},
\label{diamagnetic_operator}
\end{eqnarray}
where $H_{\gamma}$ is the Hamiltonian of the $\gamma$ degree of freedom in the normal state, and $V_{\rm cell}$ is the volume of the unit cell. The expectation value of this operator over the BCS ground state is given by 
\begin{eqnarray}
&&D_{\mu\nu}^{\,d}={\rm Tr}\left[\rho\,{\hat D}_{\mu\nu}^{\,d}\right]/{\rm Tr}\left[\rho\right]
\nonumber \\
&&=\frac{e^{2}}{V_{\rm cell}}\sum_{\bf k}\sum_{\ell=\left\{n,m\right\}}f(\varepsilon_{\ell})\langle\ell|\partial_{\mu}\partial_{\nu}H|\ell\rangle
\nonumber \\
&&=-\frac{e^{2}}{V_{\rm cell}}\sum_{\bf k}\sum_{nm}\frac{f(\varepsilon_{n})-f(\varepsilon_{m})}{\varepsilon_{n}-\varepsilon_{m}}
\nonumber \\
&&\times\left\{\langle n|\partial_{\mu}H|m\rangle\langle m|\partial_{\nu}H|n\rangle+\langle n|\partial_{\nu}H|m\rangle\langle m|\partial_{\mu}H|n\rangle\right\}
\nonumber \\
&&=-\frac{2e^{2}}{V_{\rm cell}}\sum_{\bf k}\sum_{nm}\left[f(\varepsilon_{n})-f(\varepsilon_{m})\right](\varepsilon_{n}-\varepsilon_{m})\,g_{\mu\nu}^{nm}
\label{smunu_calculation}
\end{eqnarray}
where $\rho$ is the density matrix. The detail of the derivation is given in Appendix \ref{apx:diamagnetic_current_quantum_metric}. Our formalism reveals that the diamagnetic superfluid weight at zero temperature $f(\varepsilon_{n})-f(\varepsilon_{m})=1$ is given by the momentum integration of the quasihole quantum metric $g_{\mu\nu}^{nm}$ of each pair of quasiparticle-quasihole states multiplied by the energy difference $\varepsilon_{n}-\varepsilon_{m}$, indicating the quantum geometric origin of the diamagnetic current. This feature is very similar to the dielectric and optical properties of semiconductors and insulators that are given by the momentum integration of the valence band quantum metric times an energy-dependent kernel\cite{Ahn22,Komissarov24,Cardenas24_spread_Wannier,Chen25_optical_marker}. We emphasize that this geometric interpretation of Meissner effect is only valid for $s$-wave SCs. The case of unconventional SCs where the pairing terms have momentum dependence is briefly remarked in Appendix \ref{apx:remark_unconventional_SC}.


The diagonal elements $D_{\mu\mu}^{\,d}$ are positive numbers since $\varepsilon_{n}-\varepsilon_{m}<0$ and $g_{\mu\mu}^{nm}>0$. Denoting the total current in London equations in SI units along the $\mu$ direction by 
\begin{eqnarray}
j_{d,\mu}+j_{p,\mu}\approx-\left(D_{\mu\mu}^{\,d}+D_{\mu\mu}^{\,p}\right)A^{\mu}=-\frac{1}{\mu_{0}\lambda_{L}^{2}}A^{\mu},\;\;\;
\label{London_eqs}
\end{eqnarray}
where $\mu_{0}$ is the permeability, the diamagnetic part is usually much larger than the paramagnetic part $D_{\mu\mu}^{\,d}\gg\langle {\hat D}_{\mu\mu}^{\,p}\rangle$, and we have assumed that diagonal elements dominate in practice and determine the inverse of the penetration depth square $1/\lambda_{L}^{2}$. Since the type of SC depends on the dimensionless Ginzburg-Landau parameter\cite{Ginzburg50} between penetration depth and the Ginzburg-Landau coherence length $\kappa=\lambda_{L}/\xi_{GL}$, and $\xi_{GL}$ comes from the expansion of free energy that is unrelated to the quasihole quantum geometric\cite{Gorkov59,Nagaosa99}, one sees that the type of SC is directly determined by $\left(D_{\mu\mu}^{\,d}\right)^{-1/2}$ through $\lambda_{L}$. We are lead to conclude that the type I with $\kappa<1/\sqrt{2}$ corresponds to materials with a larger while type II with $\kappa>1/\sqrt{2}$ corresponds to a smaller quantum metric times quasihole energy on average, indicating that the quantum geometry also influences the criterion for the vortex state.

\subsection{Superfluid weight marker} 

Our formalism also allows to map the superfluid weight to real space as a superfluid weight marker. This is done through utilizing the same projector formalism that maps the topological invariants to topological markers\cite{Bianco11}, and the fidelity number to a fidelity marker\cite{Marrazzo17_2,deSousa23_fidelity_marker}. Consider that the eigenstates $|E_{\ell}\rangle$ a lattice Hamiltonian $H$ of the conventional SC under question have been found via diagonalization $H|E_{\ell}\rangle=E_{\ell}|E_{\ell}\rangle$. To connect the superfluid weight to the latttice eigenstates $|E_{\ell}\rangle$, we observe that Eq.~(\ref{smunu_calculation}) may be expressed by
\begin{eqnarray}
&&D_{\mu\nu}^{\,d}
=-2\frac{e^{2}}{V_{\rm cell}}\int\frac{d^{D}{\bf k}}{(2\pi\hbar/a)^{D}}\sum_{nm}\left[f(\varepsilon_{n})-f(\varepsilon_{m})\right]
\nonumber \\
&&\;\;\;\;\;\;\;\;\times\left(\varepsilon_{n}-\varepsilon_{m}\right)\left\{\frac{1}{2}\langle\partial_{\mu}n|m\rangle\langle m|\partial_{\nu}n\rangle+(\mu\leftrightarrow\nu)\right\}
\nonumber \\
&&=-2\frac{e^{2}}{V_{\rm cell}\hbar^{2}}\int\frac{d^{D}{\bf k}}{(2\pi\hbar/a)^{D}}\sum_{nm}\left[f(\varepsilon_{n})-f(\varepsilon_{m})\right]\left(\varepsilon_{n}-\varepsilon_{m}\right)
\nonumber \\
&&\;\;\;\;\;\;\;\;\times\left\{\langle\psi_{n}^{\bf k}|{\hat\mu}|\psi_{m}^{\bf k}\rangle\langle \psi_{m}^{\bf k}|{\hat\nu}|\psi_{n}^{\bf k}\rangle+(\mu\leftrightarrow\nu)\right\}
\nonumber \\
&&=-2\frac{e^{2}}{V_{\rm cell}\hbar^{2}}\int\frac{d^{D}{\bf k}}{(2\pi\hbar/a)^{D}}\int\frac{d^{D}{\bf k'}}{(2\pi\hbar/a)^{D}}
\nonumber \\
&&\;\;\;\;\;\;\;\;\times\sum_{nm}\left[f(\varepsilon_{n})-f(\varepsilon_{m})\right]\left(\varepsilon_{n}-\varepsilon_{m}\right)
\nonumber \\
&&\;\;\;\;\;\;\;\;\times\left\{\langle\psi_{n}^{\bf k}|{\hat\mu}|\psi_{m}^{\bf k'}\rangle\langle \psi_{m}^{\bf k'}|{\hat\nu}|\psi_{n}^{\bf k}\rangle+(\mu\leftrightarrow\nu)\right\}
\nonumber \\
&&=-\frac{e^{2}}{NV_{\rm cell}\hbar^{2}}\sum_{nm}\left[\langle E_{n}|{\hat\mu}|E_{m}\rangle\langle E_{m}|{\hat\nu}|E_{n}\rangle+(\mu\leftrightarrow\nu)\right]
\nonumber \\
&&\;\;\;\;\;\;\;\;\times\left[f(E_{n})-f(E_{m})\right]\left(E_{n}-E_{m}\right)
\nonumber \\
&&=-\frac{e^{2}}{NV_{\rm cell}\hbar^{2}}\sum_{nm}{\rm Tr}\left[{\hat\mu}{\hat S}_{m}{\hat\nu}{\hat S}_{n}+(\mu\leftrightarrow\nu)\right]
\nonumber \\
&&\;\;\;\;\;\;\;\;\times\left[f(E_{n})-f(E_{m})\right]\left(E_{n}-E_{m}\right).
\label{superfluid_weight_detailed_derivation}
\end{eqnarray}
In this expression, we have used $i\langle m^{\bf k}|\partial_{\mu}n^{\bf k}\rangle=\langle\psi_{m}^{\bf k}|{\hat\mu}|\psi_{n}^{\bf k}\rangle/\hbar$ where $|\psi_{n}^{\bf k}\rangle$ is the full Bloch state and $|n^{\bf k}\rangle$ is its periodic part, and ${\hat\mu}$ is the position operator. In the last line of this expression we have introduced the projector to a specific positive eigenenergy state ${\hat S}_{m}=|E_{m}\rangle\langle E_{m}|$ and a specific negative eigenenergy state ${\hat S}_{n}=|E_{n}\rangle\langle E_{n}|$. This expression allows to convert the integration of momentum eigenstates to the lattice eigenstates $\left\{|E_{m}\rangle,|E_{n}\rangle\right\}$ of positive $E_{m}>0$ and negative $E_{n}<0$ eigenenergies of a real space lattice Hamiltonian satisfying $H|E_{\ell}\rangle=E_{\ell}|E_{\ell}\rangle$.

One may further introduce the operator 
\begin{eqnarray}
\widetilde{\cal M}_{\mu}=\sum_{nm}{\hat S}_{n}{\hat\mu}{\hat S}_{m}\sqrt{\left[f(E_{n})-f(E_{m})\right]\left(E_{n}-E_{m}\right)},\;\;\;
\end{eqnarray}
in terms of which Eq.~(\ref{superfluid_weight_detailed_derivation}) becomes
\begin{eqnarray}
&&D_{\mu\nu}^{\,d}=-\frac{e^{2}}{NV_{\rm cell}\hbar^{2}}{\rm Tr}\left[\widetilde{\cal M}_{\mu}\widetilde{\cal M}_{\nu}^{\dag}
+\widetilde{\cal M}_{\nu}\widetilde{\cal M}_{\mu}^{\dag}\right]
\nonumber \\
&&=\sum_{\bf r}D_{\mu\nu}^{\,d}({\bf r}).
\end{eqnarray}
The diagonal elements in the summation gives the local marker of superfluid weight at site ${\bf r}$
\begin{eqnarray}
&&D_{\mu\nu}^{\,d}({\bf r})=-\frac{e^{2}}{V_{\rm cell}\hbar^{2}}\sum_{\gamma}\langle{\bf r},\gamma|\left[\widetilde{\cal M}_{\mu}\widetilde{\cal M}_{\nu}^{\dag}
+\widetilde{\cal M}_{\nu}\widetilde{\cal M}_{\mu}^{\dag}\right]|{\bf r},\gamma\rangle,
\nonumber \\
\end{eqnarray}
after summing all the degrees of freedom $\gamma$ in a unit cell. Physically, the quantity $D_{\mu\nu}^{\,d}({\bf r})$ represents the diamagnetic superfluid weight at site ${\bf r}$ that promotes the Meissner effect to be locally defined
\begin{eqnarray}
j_{d,\mu}({\bf r})=D_{\mu\nu}^{\,d}({\bf r})A^{\nu}.
\end{eqnarray}
For 3D conventional SCs, the local penetration depth can be estimated as
\begin{eqnarray}
\lambda_{L}({\bf r})\approx\left[\frac{\mu_{0}}{3}\sum_{\mu}D_{\mu\mu}^{\,d}({\bf r})\right]^{-1/2},
\end{eqnarray}
assuming the paramagnetic part $\langle D_{\mu\nu}^{\,p}\rangle$ in Eq.~(\ref{London_eqs}) can be ignored. Our formalism thus allows to investigate the effect of spatial inhomogeneity on the Meissner effect, as will be demonstrated in Sec.~\ref{sec:applications_to_single_band_SC}.


At zero temperature $f(E_{n})-f(E_{m})=1$, the marker can be further simplified. In this case, we introduce the projectors 
\begin{eqnarray}
&&{\hat P}=\int\frac{d^{D}{\bf k}}{(2\pi\hbar/a)^{D}}|\psi_{n}^{\bf k}\rangle\langle\psi_{n}^{\bf k}|
=\sum_{n}|E_{n}\rangle\langle E_{n}|,
\nonumber \\
&&{\hat P}_{E}=\int\frac{d^{D}{\bf k}}{(2\pi\hbar/a)^{D}}\,\varepsilon_{n}|\psi_{n}^{\bf k}\rangle\langle\psi_{n}^{\bf k}|=\sum_{n}E_{n}|E_{n}\rangle\langle E_{n}|,
\nonumber \\
&&{\hat Q}=\int\frac{d^{D}{\bf k}}{(2\pi\hbar/a)^{D}}|\psi_{m}^{\bf k}\rangle\langle\psi_{m}^{\bf k}|
=\sum_{m}|E_{m}\rangle\langle E_{m}|,
\nonumber \\
&&{\hat Q}_{E}=\int\frac{d^{D}{\bf k}}{(2\pi\hbar/a)^{D}}\,\varepsilon_{m}|\psi_{m}^{\bf k}\rangle\langle\psi_{m}^{\bf k}|
=\sum_{m}E_{m}|E_{m}\rangle\langle E_{m}|,
\nonumber \\
\label{PE_QE_projectors}
\end{eqnarray}
in terms of which the superfluid weight at zero temperature becomes 
\begin{eqnarray}
D_{\mu\nu}^{\,d}|_{T=0}&=&-\frac{e^{2}}{NV_{\rm cell}\hbar^{2}}{\rm Tr}\left[{\hat P}_{E}{\hat \mu}{\hat Q}{\hat \nu}{\hat P}-{\hat P}{\hat \mu}{\hat Q}_{E}{\hat \nu}{\hat P}\right.
\nonumber \\
&&\;\;\;\;\;\;\;\left.+{\hat P}_{E}{\hat \nu}{\hat Q}{\hat \mu}{\hat P}-{\hat P}{\hat \nu}{\hat Q}_{E}{\hat \mu}{\hat P}\right].
\label{superfluid_weight_marker_T0}
\end{eqnarray}
Compared to the fidelity marker formalism that maps the fidelity to real space\cite{deSousa23_fidelity_marker}, the only difference in the parenthesis of Eq.~(\ref{superfluid_weight_marker_T0}) is that the projector $\left\{P_{E},Q_{E}\right\}$ in Eq.~(\ref{PE_QE_projectors}) contains an extra factor of eigenenergy $\left\{E_{n},E_{m}\right\}$ owing to the formula for ${\hat D}_{\mu\nu}^{\,d}$ in Eq.~(\ref{smunu_calculation}). As a result, the marker at zero temperature is given by
\begin{eqnarray}
D_{\mu\nu}^{\,d}({\bf r})|_{T=0}&=&-\frac{e^{2}}{V_{\rm cell}\hbar^{2}}\sum_{\gamma}\langle{\bf r},\gamma|\left[{\hat P}_{E}{\hat \mu}{\hat Q}{\hat \nu}{\hat P}-{\hat P}{\hat \mu}{\hat Q}_{E}{\hat \nu}{\hat P}\right.
\nonumber \\
&&\;\;\;\;\;\;\;\left.+{\hat P}_{E}{\hat \nu}{\hat Q}{\hat \mu}{\hat P}-{\hat P}{\hat \nu}{\hat Q}_{E}{\hat \mu}{\hat P}\right]|{\bf r},\gamma\rangle,
\label{superfluid_weight_marker_T0_atr}
\end{eqnarray}
which is a numerically very convenient tool that is readily applied to any multiband conventional SC.

\section{Applications to single-band conventional superconductors \label{sec:applications_to_single_band_SC}}

\subsection{Single-band conventional superconductors with particle-hole symmetry} 

We proceed to use single-band conventional SCs to demonstrate these points, in which case there is only one $\varepsilon_{n}$ and one $\varepsilon_{m}$, so we drop the summation $\sum_{n}$ and $\sum_{m}$ everywhere in the formalism in Sec.~\ref{sec:quantum_geometry_Meissner}. The momentum space Hamiltonian in this case can be conveniently described within the context of $2\times 2$ Dirac Hamiltonian 
\begin{eqnarray}
H({\bf k})={\bf d}\cdot{\boldsymbol\sigma}=d_{1}\sigma_{1}+d_{3}\sigma_{3},
\label{singlet_Dirac_Hamiltonian}
\end{eqnarray}
where $\sigma_{i}$ are the Pauli matrices, $d_{1}=\Delta$ is the superconducting gap, and $d_{3}=\varepsilon_{\bf k}$ is the dispersion in the normal state. The basis of the Hamiltonian is $|\psi_{\bf k}\rangle=(c_{\bf k\uparrow},c_{\bf -k\downarrow}^{\dag})^{T}$, and $d=\sqrt{d_{1}^{2}+d_{3}^{2}}=\sqrt{\varepsilon_{\bf k}^{2}+\Delta^{2}}=\varepsilon_{m}=-\varepsilon_{n}$ gives the dispersion of the two bands. The quasihole quantum metric is\cite{Porlles23_SC_metric}
\begin{eqnarray}
&&g_{\mu\nu}=\frac{\Delta^{2}}{4E_{\bf k}^{4}}v_{\mu}v_{\nu},
\label{bare_quantum_metric}
\end{eqnarray}
where $v_{\mu}=\partial_{\mu}\varepsilon_{\bf k}$ is the group velocity in the normal state. The homogeneous superfluid weight in Eq.~(\ref{smunu_calculation}) can then be calculated once a normal state dispersion $\varepsilon_{\bf k}$ is given.

For $D=2$ and $D=3$ single-band SCs treated within parabolic band approximation $\varepsilon_{\bf k}=k^{2}/2m-k_{F}^{2}/2m$, analytical results of the fidelity number ${\cal G}_{\mu\mu}$ has been given\cite{Porlles23_SC_metric}. Further assuming a cubic lattice $V_{\rm cell}=a^{D}$ with lattic constant $a$ and using ${\rm Tr}{\cal G}_{\mu\nu}=D\times{\cal G}_{\mu\mu}$, the gauge-invariant part $\Omega_{I}$ of the spread of quasihole Wannier function is
\begin{eqnarray}
\Omega_{I}=a^{2}\frac{\pi^{2}}{c_{D}\sqrt{2}}\left(\frac{\xi_{0}}{a}\right)\left(\frac{k_{F}}{2\pi\hbar/a}\right)^{D-1},
\end{eqnarray}
with the coefficient $c_{D}=2$ at $D=3$ and $c_{D}=4$ at $D=2$. Thus we see that the spread in units of lattice constant square $a^{2}$ is determined by the ratio between coherence length $\xi_{0}=\hbar v_{F}/\pi\Delta$ and the lattice constant $a$ multiplied by the ratio between Fermi momentum $k_{F}$ and BZ boundary $2\pi\hbar/a$. The dependence on the coherence length implies that a larger gap and a smaller Fermi velocity (or equivalently a smaller normal state band width) will have a very small $\Omega_{I}$, i.e., a very localized quasihole Wannier function, indicating a very stable SC state. In contrast, a large coherence length implies a very extended quasihole Wannier function that is more mobile and easier to excite, and hence a relatively unstable SC state.

\subsection{2D superconductor on a square lattice}

As a concrete example for the superfluid weight marker, we examine the mean field theory of 2D $s$-wave SC on a square lattice described by the lattice Hamiltonian
\begin{eqnarray}
H&=&\sum_{\langle ij\rangle\sigma}t\,c_{i\sigma}^{\dag}c_{j\sigma}-\sum_{i\sigma}\mu\,c_{i\sigma}^{\dag}c_{i\sigma}
+\sum_{i\in{\rm imp}}U_{imp}\,c_{i\sigma}^{\dag}c_{i\sigma}
\nonumber \\
&&+\sum_{i}\left(\Delta_{i}c_{i\uparrow}^{\dag}c_{i\downarrow}^{\dag}
+\Delta_{i}^{\ast}c_{i\downarrow}c_{i\uparrow}\right),
\label{mean_field_SC_Hamiltonian}
\end{eqnarray}
where $c_{i\sigma}$ is the electron annihilation operator at site $i$ with spin $\sigma$, $t=-1$ is the nearest-neighbor hopping that serves as energy unit, $\mu=-0.2$ is the chemical potential, $\Delta_{i}$ is the on-site $s$-wave pairing to be determined self-consistently, and $U_{imp}$ is the nonmagnetic impurity potential at site $i$. We diagonalize the mean field Hamiltonian $H={\rm const.}+\sum_{\bf k\alpha}E_{\ell}\gamma_{\ell\alpha}^{\dag}\gamma_{\ell\alpha}$ by a Bogoliubov transformation
\begin{eqnarray}
&&c_{i\uparrow}=\sum_{\ell}\gamma_{\ell\uparrow}u_{\ell}(i)-\gamma_{\ell\downarrow}^{\dag}v_{\ell}^{\ast}(i),
\nonumber \\
&&c_{i\downarrow}=\sum_{\ell}\gamma_{\ell\downarrow}u_{\ell}(i)+\gamma_{\ell\uparrow}^{\dag}v_{\ell}^{\ast}(i),
\end{eqnarray}
where $\gamma_{\ell\sigma}$ is the annihilation operator of the Bogoliubov quasiparticles. The wave functions $\left\{u_{\ell}(i),v_{\ell}(i)\right\}$ and eigenenergy $E_{\ell}$ satisfy
\begin{eqnarray}
E_{\ell}u_{\ell}(i)&=&\sum_{\langle ij\rangle}t\,u_{\ell}(j)-\mu u_{\ell}(j)
\nonumber \\
&&+U_{imp}\delta_{i\in imp}u_{\ell}(i)+\Delta_{i}v_{\ell}(i),
\nonumber \\
E_{\ell}v_{\ell}(i)&=&-\sum_{\langle ij\rangle}t\,v_{\ell}(j)
+\mu v_{\ell}(j)
\nonumber \\
&&-U_{imp}\delta_{i\in imp}v_{\ell}(i)+\Delta_{i}^{\ast}u_{\ell}(i).
\label{BdG_u_v_equations}
\end{eqnarray}
The pairing amplitude at site $i$ is then detemrined by
\begin{eqnarray}
\Delta_{i}=\sum_{\ell}V\left[2f(E_{\ell})-1\right]u_{\ell}(i)v_{\ell}^{\ast}(i),
\label{BdG_gap_eq}
\end{eqnarray}
where $f(E_{\ell})=(e^{E_{\ell}/k_{B}T}+1)^{-1}$ is the Fermi distribution and $V<0$ is the pairing interaction. Equations (\ref{BdG_u_v_equations}) and (\ref{BdG_gap_eq}) are solved self-consistently until the local pairing amplitude $\Delta_{i}$ converges at a given pairing interaction $V$, and for simplicity we focus on zero temperature $T=0$. The resulting quasihole states $|E_{n}\rangle$ with $E_{n<0}$ and quasiparticle states $|E_{m}\rangle$ with $E_{m}>0$ are then inserted into the projectors $P$ and $Q$ in Eq.~(\ref{superfluid_weight_marker_T0_atr}) to calculate the zero-temperature superfluid weight marker. We will limit our discussion to the weak magnetic field limit ${\bf B}={\boldsymbol\nabla}\times{\bf A}\rightarrow 0$ such that the vector potential is relatively uniform in the nanometer scale. For concreteness, a uniform vector potential pointing along $-{\hat{\bf x}}$ direction is assumed, so the diamagnetic current described by the superfluid weight marker as a two-component vector field $(D_{xx}^{d}({\bf r}),D_{yx}^{d}({\bf r}))$ should predominantly point at the $+{\hat{\bf x}}$ direction. Moreover, by summing the $x$-component of the marker $D_{xx}^{d}(x,y)$ over the transverse coordinate $y$ at a given longitudinal position $x$, we can define an average flux flowing across the cross section at $x$
\begin{eqnarray}
f_{xx}^{d}(x)=\frac{1}{N_{y}}\sum_{y}D_{xx}^{d}(x,y),
\end{eqnarray}
where $N_{y}$ is the number of sites of the cross section, which helps to clarify whether the average flux of diamagnetic current remains conserved near the impurity site.

\begin{figure}[ht]
\begin{center}
\includegraphics[clip=true,width=0.99\columnwidth]{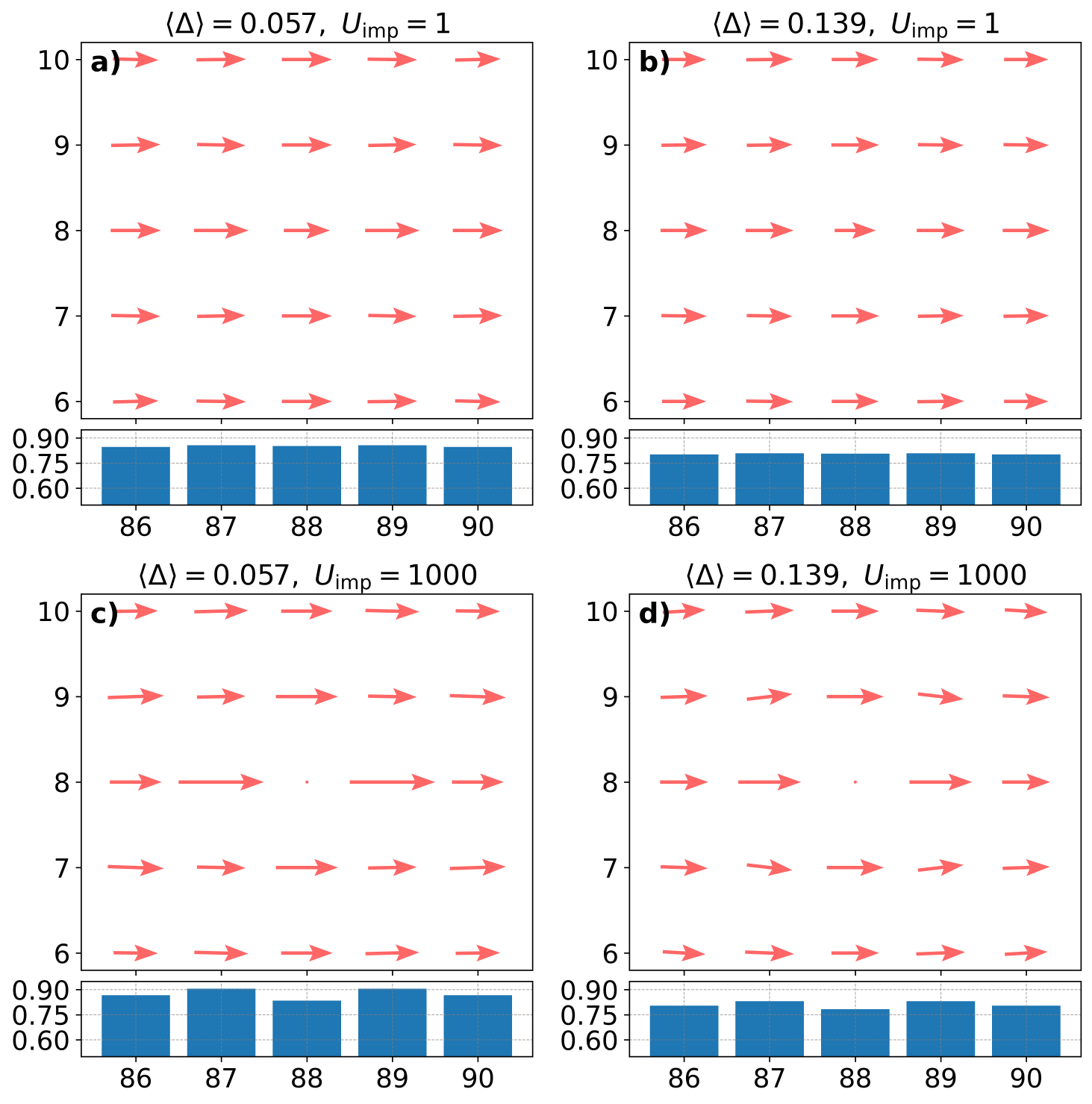}
\caption{The diamagnetic current (red arrows) described by the vector field $(D_{xx}^{d},D_{yx}^{d})$ of superfluid weight marker in a 2D $s$-wave SC in the presence of a single nonmagnetic impurity, simulated on a $175\times 15$ lattice and assuming a vector potential ${\bf A}$ pointing to the $-{\hat{\bf x}}$ direction. The average flux $f_{xx}^{d}$ (blue bars) of diamagnetic current flowing through the cross section at $x$ is given in the lower part of each panel. We examine the average SC gap $\langle\Delta\rangle$ and a single impurity located at the center with impurity potential $U_{imp}$ taking the values indicated by in each panel. } 
\label{fig:2DSSC_single_imp}
\end{center}
\end{figure}

In Fig.~\ref{fig:2DSSC_single_imp}, we present the numerical results for the vector field $(D_{xx}^{d}({\bf r}),D_{yx}^{d}({\bf r}))$ and the average flux $f_{xx}^{d}(x)$ at $x$ in the presence of a single impurity, solved at pairing interactions $V=-1.1$ and $-1.4$. A rectangular lattice of size $175\times 15$ is chosen for simulation because it is known that in the topological marker formalism with a straightforward assignment of the position operator ${\hat\mu}={\rm diag}(1,2,3...)\otimes\gamma$, the marker on the boundary sites becomes inaccurate, so we choose to elongate one direction and plot only the center sites to avoid the error. For the weak impurity potential cases $U_{imp}=1$ shown in Fig.~\ref{fig:2DSSC_single_imp} (a) and (c), the vector field indicates that the diamagnetic current is predominantly along $+{\hat{\bf x}}$ direction, as we argued above, and the flow is only weakly perturbed by the impurity with an average flux $f_{xx}^{d}(x)$ roughly conserved. On the other hand, for the large impurity potential $U_{imp}=1000$ that effectively projects out the impurity site, the flow on the impurity site is completely suppressed, and the flow near the impurity site displays a pattern that seems to circumvent the impurity site in a way that the average flux $f_{xx}^{d}(x)$ is not conserved. This feature occurs at both large and small SC gap, respectively, but the circumvention is more obvious at large impurity potential.


\begin{figure}[ht]
\begin{center}
\includegraphics[clip=true,width=0.99\columnwidth]{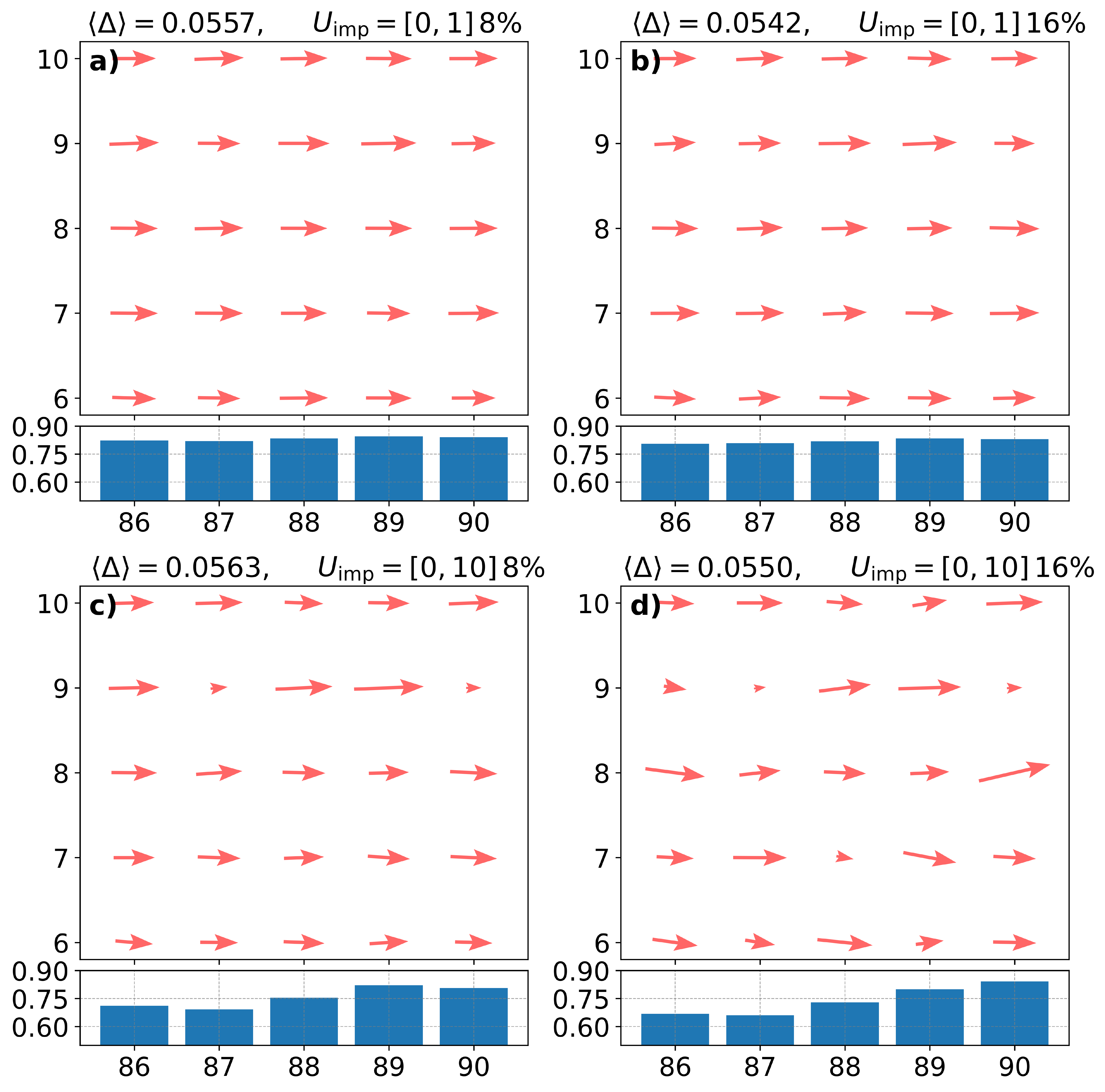}
\caption{The diamagnetic current (red arrows) and average flux (blue bars) in a 2D $s$-wave SC in the presence of random nonmagnetic impurities, plotted in the center region of a $175\times 15$ lattice with a vector potential ${\bf A}$ pointing to the $-{\hat{\bf x}}$ direction. We examine impurities with density $n_{imp}$, and each impurity produces an impurity potential randomly distributed within an interval. The average SC gap $\langle\Delta\rangle$ and impurity parameters in each panel are indicated in each panel.} 
\label{fig:2DSSC_many_imp}
\end{center}
\end{figure}

We further examine the effect of many impurities by introducing multiple impurities of density $n_{imp}$ into the system, and each impurity is assigned with an impurity potential $U_{imp}$ randomly distributed within an interval. The resulting vector field $(D_{xx}^{d}({\bf r}),D_{yx}^{d}({\bf r}))$ and average flux $f_{xx}^{d}(x)$ through the cross section at $x$ are shown in Fig.~\ref{fig:2DSSC_many_imp}. The results indicate that at small impurity density $n_{imp}=8\%$ and small impurity potential interval $\left[0,1\right]$, the diamagnetic current remains predominantly pointing in the direction $+{\hat{\bf x}}$ opposite to the vector potential with a roughly uniform magnitude, suggesting that the superfluid weight and penetration depth are not much influenced by weak impurities. However, as the impurity density increases to $n_{imp}=16\%$ and the impurity potential interval increases to $\left[0,10\right]$, the diamagnetic current starts to become turbulent in a way that reminisces the circumventing current in the single-impurity case shown in Fig.~\ref{fig:2DSSC_single_imp}. In addition, the overall magnitude of the vector field $(D_{xx}^{d}({\bf r}),D_{yx}^{d}({\bf r}))$ is reduced, resulting in a reduction of average flux $f_{xx}^{d}(x)$ through the cross section everywhere in the system, indicating the the superfluid weight in the macroscopic scale is suppressed by strong impurities. Our result thus suggests that strong nonmagnetic impurities have detrimental effects on the diamagnetic current on 2D $s$-wave SCs, and our superfluid weight marker serves as a useful tool to quantify their influence in the atomic scale. Finally, we anticipate that the reduction of diamagnetic current on the impurity site may be detected in 2D SCs by microscopes that can resolve magnetic field in the atomic scale, such as the Magnetic Exchange Force Microscopy (MExFM)\cite{Kaiser07}, although it remains unclear whether such technique can be applied to superconductors.


\begin{figure}[ht]
\begin{center}
\includegraphics[clip=true,width=0.99\columnwidth]{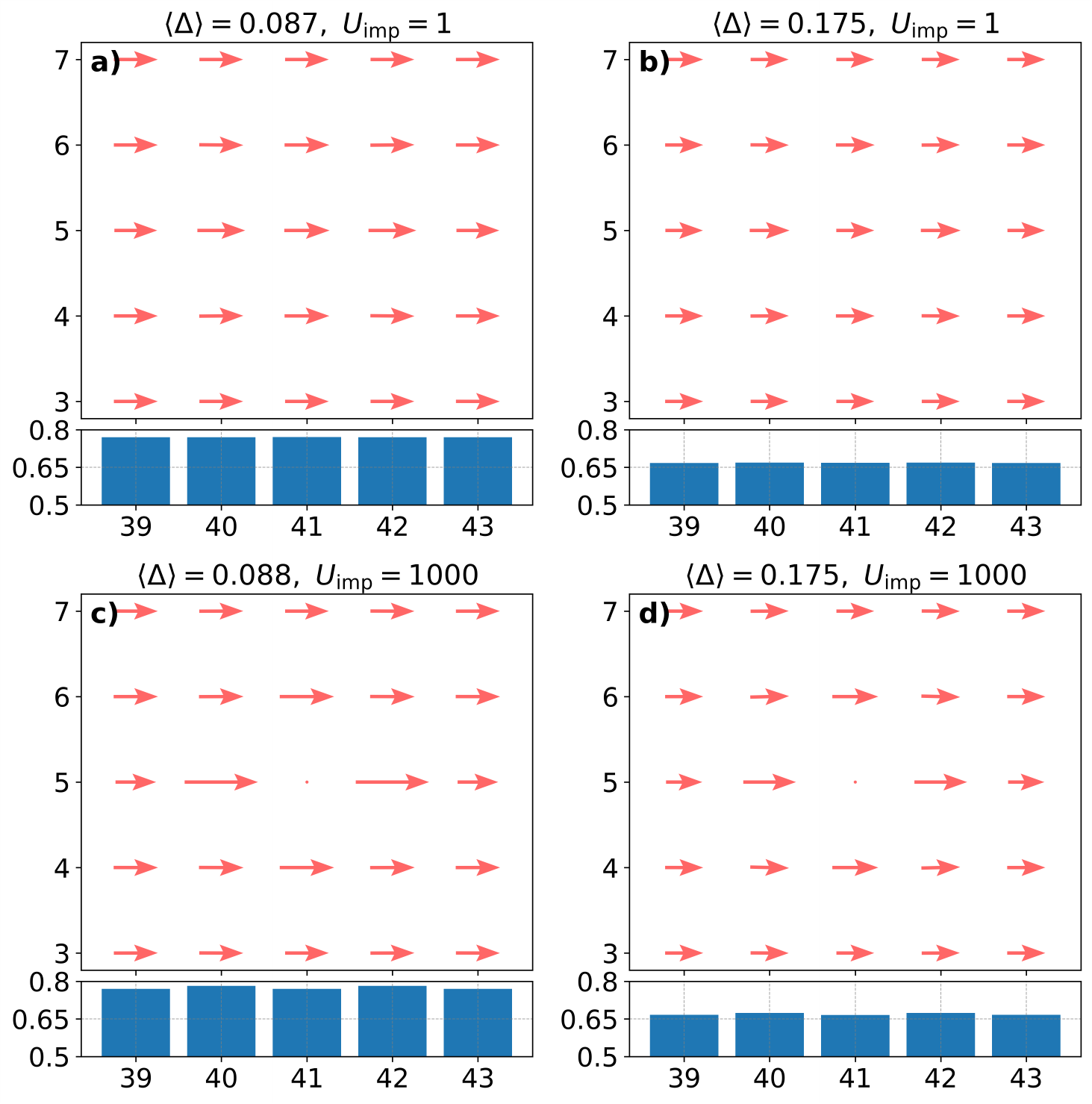}
\caption{The diamagnetic current described by the planar component of the vector field $(D_{xx}^{d},D_{yx}^{d})$ of superfluid weight marker in a 3D $s$-wave SC in the presence of a single nonmagnetic impurity, assuming a vector potential ${\bf A}$ pointing to the $-{\hat{\bf x}}$ direction. We plot the flow pattern in the $xy$-plane cutting through and near the impurity site in a $81\times 9\times 9$ lattice. The average flux $f_{xx}^{d}$ flowing through the cross section at $x$ is given in the lower part of each panel. The values of average SC gap $\langle\Delta\rangle$ and impurity potential $U_{imp}$ are indicated by in each panel.} 
\label{fig:3DSSC_single_imp_fig}
\end{center}
\end{figure}

\begin{figure}[ht]
\begin{center}
\includegraphics[clip=true,width=0.99\columnwidth]{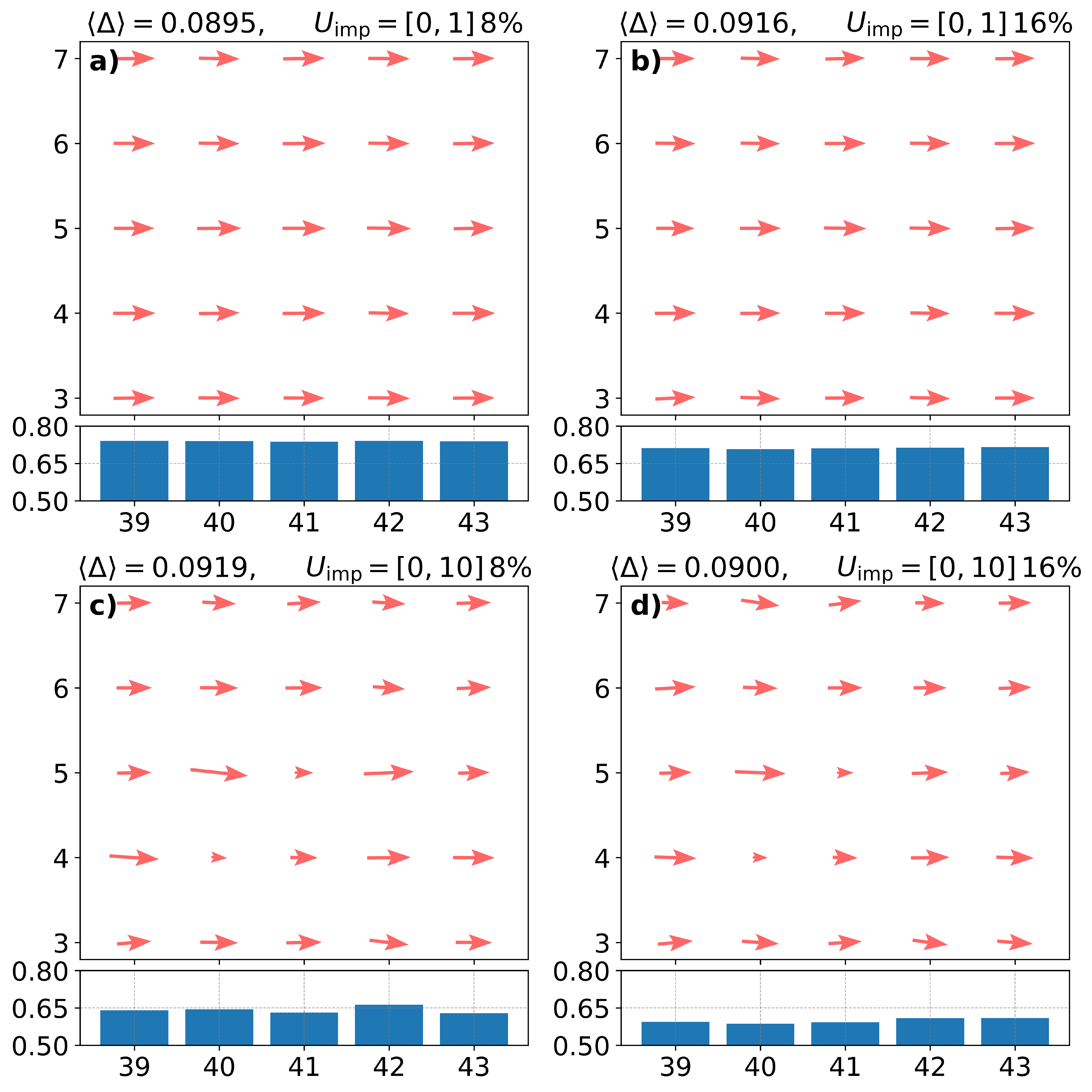}
\caption{The diamagnetic current in a 3D $s$-wave SC in the presence of random nonmagnetic impurities and a vector potential ${\bf A}$ pointing to the $-{\hat{\bf x}}$ direction, plotted for the $xy$-plane cutting through the center of an $81\times 9\times 9$ lattice. The average flux $f_{xx}^{d}$ flowing through the cross section at $x$ is given in the lower part of each panel. We examine impurities with density $n_{imp}$, and each impurity produces an impurity potential randomly distributed within an interval, with values indicated in each panel.} 
\label{fig:3DSSC_many_imp_fig}
\end{center}
\end{figure}

\subsection{3D superconductor on a cubic lattice}

We proceed to examine 3D $s$-wave SCs using the same lattice model of Eq.~(\ref{mean_field_SC_Hamiltonian}) but defined on a cubic lattice, which is more relevant to realistic $s$-wave SCs. In Fig.~\ref{fig:3DSSC_single_imp_fig}, we show the planar component $(D_{xx}^{d},D_{yx}^{d})$ of the superfluid weight marker on the $xy$-plane cutting though a single impurity, solved at pairing interactions $V=-1.6$ and $-1.9$. One sees a similar feature as the 2D results presented in Fig.~\ref{fig:2DSSC_single_imp}, namely the superfluid weight is suppressed on the impurity site, especially at large impurity potential. The diamagnetic current also shows the feature of circumventing the impurity site when the impurity potential is large, although this feature is more moderate compared to 2D. In the presence of many impurities, the diamagnetic current shown in Fig.~\ref{fig:3DSSC_many_imp_fig} remains predominantly flowing along $+\hat{\bf x}$ direction, with a turbulent pattern and reduced magnitude, causing the average flux $f_{xx}^{d}$ to reduce just like in 2D. Consequently, the penetration length in the macroscopic scale is increased by nonmagnetic impurities according to the London equations in Eq.~(\ref{London_eqs}). This conclusion is entirely consistent with the seminal experimental result of Pippard that shows the increase of penetration depth in Sn upon reducing the normal state mean free path\cite{Pippard53}. Our calculation thus offers a microscopic explanation to this experimental result in terms of self-consistently solving the BdG equations and the superfluid weight marker, and moreover delineates how the diamagnetic current is influenced by the impurities in the atomic scale. Finally, we remark that in highly disordered SCs, the paramagnetic current caused by the $e{\hat v}_{\mu}A^{\mu}$ term in Eq.~(\ref{Hamiltonian_expansion_vector_potential}) may contribute significantly and diminish the Meissner effect, together with other diffusive effects beyond the perturbative expansion of Eq.~(\ref{Hamiltonian_expansion_vector_potential}). Currently it remains unclear to us if the paramagnetic current, usually calculated by linear response theory\cite{Liang17}, can also be mapped to real space as a marker and compare with our superfluid weight marker to give a more complete description of Meissner effect under the influence of strong disorder. Further investigations are necessary to clarify this highly disordered limit.

\section{Conclusions}

In summary, we clarify the quantum geometric origins of a number of defining properties of conventional multiband superconductors with arbitrary intraband and interband $s$-wave pairings. We first elaborate that the momentum-integration of the quasihole quantum metric is equal to the gauge-invariant part of a spread of quasihole Wannier function, proposed to characterize the stability of the SC state. Turning to the Meissner effect, we show that the superfluid weight in London equations is given by the momentum-integration of elements of quantum metric times the quasiparticle energy, pointing to the quantum geometric origin of Meissner effect and vortex state. A remarkable consequence of this origin is that it allow to map the superfluid weight to individual lattice sites as a superfluid weight marker, enabling the effect of disorder on diamagnetic current to be investigated self-consistently through solving the BdG equations. Applying this marker to single-band SCs in 2D and 3D reveals a diamagnetic current that circumvents individual impurities and becomes turbulent when multiple impurities are present, and reproduces the experimental result of enlarged penetration depth caused by disorder. We anticipate that our superfluid weight marker can be ubiquitously applied to investigate other kinds of disorder like magnetic impurities, as well as studying other phases of SCs such as the vortex state, and may also contribute significantly to the diamagnetic current in unconventional SCs. These intriguing issues remain to be further explored.

\acknowledgements

We acknowledge the financial suppoort from the fellowship for productivity in research from CNPq.

\appendix

\section{Expression of diamagnetic current in terms of quasihole quantum metric \label{apx:diamagnetic_current_quantum_metric}}

We now detail the expression of diamagnetic current in terms of quasihole quantum metric given in Eq.~(\ref{smunu_calculation}). We first split the equation into the summation over quasiparticle $|m\rangle$ and quasihole $|n\rangle$ states
\begin{eqnarray}
D_{\mu\nu}^{\,d}&=&\frac{e^{2}}{V_{\rm cell}}\sum_{n}f(\varepsilon_{n})\langle n|\partial_{\mu}\partial_{\nu}H|n\rangle
\nonumber \\
&&+\frac{e^{2}}{V_{\rm cell}}\sum_{m}f(\varepsilon_{m})\langle m|\partial_{\mu}\partial_{\nu}H|m\rangle.
\label{superfluid_weight_derivation_meddle_step}
\end{eqnarray}
Consider the summation over quasihole states first. Using $\langle n|\partial_{\nu}H|n\rangle=\partial_{\nu}E_{n}$ and $\sum_{n'}|n'\rangle\langle n'|+\sum_{m'}|m'\rangle\langle m'|=I$, we notice that 
\begin{widetext}
\begin{eqnarray}
&&\langle n|\partial_{\mu}\partial_{\nu}H|n\rangle=\partial_{\mu}\partial_{\nu}E_{n}
-\langle\partial_{\mu} n|\partial_{\nu}H|n\rangle-\langle n|\partial_{\nu}H|\partial_{\mu}n\rangle
\nonumber \\
&&=\partial_{\mu}\partial_{\nu}E_{n}-\sum_{n'}\langle\partial_{\mu} n|n'\rangle\langle n'|\partial_{\nu}H|n\rangle-\sum_{m'}\langle\partial_{\mu} n|m'\rangle\langle m'|\partial_{\nu}H|n\rangle
-\sum_{n'}\langle n|\partial_{\nu}H|n'\rangle\langle n'|\partial_{\mu}n\rangle-\sum_{m'}\langle n|\partial_{\nu}H|m'\rangle\langle m'|\partial_{\mu}n\rangle
\nonumber \\
&&=\partial_{\mu}\partial_{\nu}E_{n}-\sum_{m'}\langle\partial_{\mu} n|m'\rangle\langle m'|\partial_{\nu}H|n\rangle-\sum_{m'}\langle n|\partial_{\nu}H|m'\rangle\langle m'|\partial_{\mu}n\rangle
\nonumber \\
&&=\partial_{\mu}\partial_{\nu}E_{n}-\sum_{m}\frac{\langle n|\partial_{\mu}H|m\rangle\langle m|\partial_{\nu}H|n\rangle}{E_{n}-E_{m}}-\sum_{m}\frac{\langle n|\partial_{\nu}H|m\rangle\langle m|\partial_{\mu}H|n\rangle}{E_{n}-E_{m}},\;\;\;\;\;
\end{eqnarray}
\end{widetext}
The $\partial_{\mu}\partial_{\nu}E_{n}$ terms drops out after momentum integration, and the kernel in the summation over quasiparticle states $\langle m|\partial_{\mu}\partial_{n}H|m\rangle$ can be manipulated in the same way. Putting these results back to Eq.~(\ref{superfluid_weight_derivation_meddle_step}) yields the final expression in Eq.~(\ref{smunu_calculation}).

We remark that the paramagnetic current derived by Liang et al.~contains matrix elements like $\langle n|\partial_{\mu}H\gamma^{z}|m\rangle$ where $\gamma^{z}$ is the Pauli matrix in the particle-hole space\cite{Liang17}. Owing to this extra $\gamma^{z}$, it is in general not possible to convert the matrix element to $\langle\partial_{\mu}n|m\rangle$ and subsequently to a local marker. Thus it remains unclear to us whether the paramagnetic current can also be converted into a marker and directly calculated from a lattice Hamiltonian.

\section{Remarks on unconventional superconductors \label{apx:remark_unconventional_SC}}

Although we have focused on conventional multiband SCs, it is entirely obvious that the quasihole quantum metric formulated in Sec.~\ref{sec:quantum_metric_multiband} also exists in unconventional SCs with other pairing symmetries, such as $p$-wave or $d$-wave, be it singlet or triplet. Although this generalization is straightforward, we remark that one must be cautious to attribute the Meissner effect in unconventional SCs to the quantum metric for the following reason. Denoting the full Hamiltonian as $H=H_{0}+H_{\Delta}$, where $H_{0}$ represents the normal state Hamiltonian and $H_{\Delta}$ the pairing terms, one notices that all three quantities $\left\{H,H_{0},H_{\Delta}\right\}$ depend on momentum ${\bf k}$ in unconventional SCs. The quantum metric derived in Sec.~\ref{sec:origin_Meissner_effect} and Appendix \ref{apx:diamagnetic_current_quantum_metric} involves the derivatives on the full Hamiltonian $H$, and hence contains the derivative on both $H_{0}$ and $H_{\Delta}$
\begin{eqnarray}
\partial_{\mu}\partial_{\nu}H=\partial_{\mu}\partial_{\nu}H_{0}+\partial_{\mu}\partial_{\nu}H_{\Delta}.
\end{eqnarray}
On the other hand, the Meissner effect is only determined by the normal state Hamiltonian $H_{0}$ due to the minimal coupling between electrons and the vector potential, as implied by Eq.~(\ref{diamagnetic_operator}). For conventional SCs, $\partial_{\mu}\partial_{\nu}H_{\Delta}=0$ vanishes since the pairing terms are all momentum-independent, and hence $\partial_{\mu}\partial_{\nu}H=\partial_{\mu}\partial_{\nu}H_{0}$, leading to the conclusion that the quantum metric directly determines the superfluid weight. In contrast, for unconventional SCs where $\partial_{\mu}\partial_{\nu}H_{\Delta}\neq 0$ due to momentum-depend pairing, we see that $\partial_{\mu}\partial_{\nu}H\neq\partial_{\mu}\partial_{\nu}H_{0}$, and hence the quantum metric does not completely determine the Meissner effect since the former contains the effect of $\partial_{\mu}\partial_{\nu}H_{\Delta}$. In this case, one may express the superfluid weight calculated from $\partial_{\mu}\partial_{\nu}H_{0}$ in terms of the quantum metric calculated from $\partial_{\mu}\partial_{\nu}H$ minus that contributed from the momentum-dependent pairing $\partial_{\mu}\partial_{\nu}H_{\Delta}$, but it is unclear to us whether the quantum metric will remain the dominant contribution. This issue has to be clarified case by case according to the unconventional SC given at hand. 


\bibliography{Literatur_abbreviated}

\end{document}